\documentclass{aa}
\usepackage{txfonts}
\usepackage{graphicx}

\begin{document}

\title{Long period variables and mass loss in the globular clusters NGC~362 and NGC~2808}

\author{
T. Lebzelter\inst{1} 
\and P.R. Wood\inst{2}
}

\institute{
Department of Astronomy, University of Vienna, Tuerkenschanzstrasse 17, A1180 Vienna, Austria
\and
Research School of Astronomy and Astrophysics, The Australian National University, Canberra, Australia}

\date{Received / Accepted}

\abstract{The pulsation periods of long period variables (LPVs) depend on
their mass and helium abundance as well as on their luminosity and metal abundance.
Comparison of the observed periods of LPVs in globular clusters
with models is capable of revealing the amount of mass lost on the giant branch
and the helium abundance.}
{We aim to determine the amount of mass loss that has occurred on the giant branches
of the low metallicity globular clusters NGC~362 and NGC~2808.  We also aim
to see if the LPVs in NGC~2808 can tell us about helium abundance variations in
this cluster.}
{We have used optical monitoring of NGC~362 and NGC~2808 to determine periods for
the LPVs in these clusters.  We have made linear pulsation models for
the pulsating stars in these clusters taking into account variations in mass
and helium abundance.}
{Reliable periods have been determined for 11 LPVs in NGC~362 and 15 LPVs in NGC~2808.
Comparison of the observed variables with models in the $\log P$--$K$ diagram
shows that mass loss of $\sim$0.15--0.2 $M_{\odot}$ is required on
the first giant branch in these clusters, in agreement with estimates from
other methods.  In NGC~2808, there is evidence that a high helium
abundance of Y\,$\sim$\,0.4 is required to explain the periods of
several of the LPVs.}
{It would be interesting to determine periods for LPVs in other Galactic
globular clusters where a helium abundance variation is suspected
to see if the completely independent test for a high helium abundance 
provided by the LPVs can confirm the high helium abundance estimates.}

\keywords{stars: AGB and post-AGB - stars: variables - Globular clusters: NGC~362, NGC~2808}

\maketitle

\section{Introduction}
Long period variability is a typical characteristic of low and
intermediate mass stars during their evolution on the Asymptotic Giant
Branch (AGB). This evolutionary phase is of high importance as it is
accompanied by high mass loss rates effectively removing the outer
layers of a star and returning its material to the interstellar
medium. Knowledge of this mass loss process has a strong impact on our
understanding of stellar and galactochemical evolution.

Mass loss and variability seem to be linked on the AGB, with stars having
high mass loss rates also showing large-amplitude pulsation. The radial
pulsation lifts substantial amounts of material to the grain formation
radius after which the radiation pressure force acting on the grains
can drive a stellar wind (e.g. Wood \cite{wood79}; H\"ofner
\cite{Hoefner08}). While this scenario is widely accepted nowadays,
the details of the mass loss process are not totally understood
yet. This is partly due to the fact that data on stars with a well
determined mass and metallicity are rare. Consequently, the dependency
of mass loss on these two parameters could not be studied sufficiently
in the past.

While it is very difficult to derive these basic parameters for field
stars, the rather homogeneous data sets provided by globular clusters
offer the possibility to work on this problem.  We can safely assume
that all stars we see at a given moment on the AGB of a cluster do not
differ significantly in their mass and age or in their
metallicity (although recent studies indicate that more than one
episode of star formation may have occured in some globular clusters
of the Milky Way, e.g. D'Antona \& Ventura \cite{DAV08}).  Therefore
the upper giant branches of globular clusters are an excellent testbed
to study the properties of evolved stars.

Wood et al. (\cite{Wood99}) have shown that long period variables
(LPVs) form a number of parallel sequences in the period-luminosity
diagram and that some of these sequences can be explained by radial
pulsation in the fundamental mode and in low order overtone modes (see
Soszy{\'n}ski et al. \cite{sosz09} and Wood \& Arnett \cite{wa10} for
recent observational and theoretical versions of the period luminosity
sequences for LPVs).  Once a star starts to lose mass at a high rate,
its total mass will start to decrease significantly, which will affect
the observed period--luminosity sequences. Lebzelter \& Wood
(\cite{LW05}) measured the $\log P$--$K$-sequences for the LPVs in the
globular cluster 47~Tuc and they found that the shape of these
sequences can only be explained if a steady decrease of mass on the
RGB and the AGB is assumed. Their paper illustrated the possibility to
investigate the mass loss rate of low mass stars by the study of an
ensemble of AGB variables in a single stellar population.

The aim of the present paper is to extend this study to two further
clusters with a metallicity different from that of 47~Tuc. The two
clusters are NGC~362 and NGC~2808. In the following we will give a
brief review of the literature on these two targets. Then we describe
the observations obtained for this study in order to make a detailed
search for LPVs.  The number of previously known LPVs is small and not
sufficient to define the path of $\log P$--$K$ sequences. Finally, we present
the $\log P$--$K$ diagram for the two clusters and discuss it in the light of
linear pulsation models and a Reimers-like mass loss rate.

\subsection{NGC~362}
With the publication of the first colour--magnitude diagram of the cluster by Menzies (\cite{Menzies67}), it
became obvious that NGC~362 does not follow the standard picture of the cluster metallicity
vs. colour--magnitude diagram (CMD) morphology. 
The horizontal branch, which is populated mainly on the red side of the instability
strip, would suggest a comparatively high metallicity (according to
Carretta \& Gratton \cite{CG97} similar to 47~Tuc, which would mean [Fe/H]=$-$0.66), while the slope of the giant
branch, the integrated colours and the cluster's spectral type indicate a lower metallicity (e.g.
Alcaino \cite{Alcaino76}; Harris \cite{Harris82}). The discrepancy appearing for this cluster provided
a major input for the discussion of the 2nd parameter problem in globular clusters (e.g. Alcaino \cite{Alcaino76}).

Several authors have published abundance studies for individual stars in NGC~362. Pilachowski (\cite{Pila81})
derived a mean [Fe/H]=$-0.9$ from several cluster giants. Gratton (\cite{Gratton87}) measured a value
of [Fe/H]=$-1.2$ from one star. 
Caldwell \& Dickens (\cite{CD86}) find [Fe/H]$=-$0.9 from 2 red giants.
Shetrone \& Keane (\cite{SK00}) derive [Fe/H]$=-$1.33$\pm$0.01 from 12 giants in agreement
with the work of Rutledge et al. (\cite{Rutledge97}) giving $-$1.36. The compilation of Harris
(\cite{Harris96}) lists $-$1.16 based on earlier literature data.
Photometric studies revealed a split of the cluster subgiants, RGB and AGB stars into a CN-rich and a 
CN-poor group, with the majority being CN-rich (Smith \cite{Smith83}, \cite{Smith84}). 
As a cause for this bimodality, Kayser et al. (\cite{KHGW08}) suggest an enrichment by 
intermediate mass AGB stars from a first generation of stars in the cluster (see also
the summary on NGC~2808 given in the next section).
The width of the
main sequence, however, indicates that there is only a small spread in the overall metal abundance
(Bolte \cite{Bolte87}). 

NGC~362 is thought to be slightly younger than other globular
clusters, in particular it appears to be 1-3 Gyr younger than NGC 288,
with which it is often compared due to the similar metallicity
(e.g. Green \& Norris \cite{GN90}; Catelan \& de Freitas Pacheco
\cite{CFP99}). Gratton et al. (\cite{Gratton97}) give an age of 7.8 to
9.3 Gyr for this cluster while Meissner \& Weiss (\cite{MW06}) find a
an age of approximately 8.5$\pm$2 Gyr.

\subsection{NGC~2808}
NGC~2808 is famous for various outstanding features in its
CMD (Walker \cite{Walker99}; Bedin et al.\,\cite{Bedin00}). Three main sequences have been identified
(Piotto et al.  \cite{Piotto07}) which are attributed to three phases
of star formation. The findings do not allow for an appreciable age
difference between the three phases. In this context, the cluster's
horizontal branch morphology is also quite puzzling, with a
combination of a red clump-like structure and long tail blueward of
the instability strip (Bedin et al. \cite{Bedin00}; D'Antona \& Caloi
\cite{dAC08}; Dalessandro et al. \cite{dal10}). The number of RR\,Lyr
stars is rather small (e.g. Clement \& Hazen \cite{CH89}; Corwin et
al.\,\cite{corwin04}).

In an attempt to explain these characteristics, it has been suggested
that gas from which the second and third generation of stars formed
has been contaminated by intermediate-mass AGB stars (D'Antona \&
Caloi \cite{dAC04}; \cite{dAC08}). NGC~2808 is one of the clusters
suspected to be massive enough to retain the AGB ejecta, thus He
enriched material would be available for star formation.  There is no
spread in the iron abundance in the cluster, hence a He abundance
spread from Y $\sim$ 0.25 to 0.4 seems necessary to explain the
triple main-sequence (MS) (Piotto et al. \cite{Piotto07}).  A further observation
supporting this scenario comes from abundance measurements of red
giants (Carretta et al. \cite{Carretta06}) which show three groups of
stars according to the O-abundance.

The alternative scenario trying to explain the HB morphology assumes
an enhanced mass loss for a fraction of the red giants before reaching
the tip of the red giant branch (RGB). Sandquist \& Martel (\cite{SM07}) found from the
comparison of cumulative observed and synthetic luminosity functions
that NGC~2808 shows a lack of stars immediately below the RGB-tip,
which they explain with this scenario.

Ferraro et al. (\cite{Ferraro90}) give a summary on literature values
for the metallicity and the reddening which range for [Fe/H] from
$-$0.96 to $-$1.48, and for E(B-V) from 0.21 to 0.34 mag,
respectively. Rutledge et al. (\cite{Rutledge97}) find [Fe/H]\,=\,$-$1.34
from spectroscopic observations of individual cluster
members. Carretta et al. (\cite{Carretta04}), also from the analysis
of individual stars, derive [Fe/H]\,=\,$-$1.14$\pm$0.01 on the Carretta \&
Gratton scale corresponding to about $-$1.27 on the Zinn \& West
scale. The compilation of Harris (\cite{Harris96}) lists
[Fe/H]\,=\,$-$1.37.

NGC~2808 is an old cluster with an age similar to 47~Tuc (Santos \&
Piatti \cite{SP04}). Quantitative age estimates scatter between 11 Gyr
(D'Antona \& Caloi \cite{dAC08}) and 12.5 to 13 Gyr (Piotto et al.
\cite{Piotto07}; D'Antona \& Caloi \cite{dAC04}).  Somewhat younger is
the determination by Meissner \& Weiss (\cite{MW06}) who give an age
as low as 8 to 10 Gyr.  In their detailed examination of globular
cluster relative ages, Marin-Franch et al. (\cite{MF09}) found that
NGC~2808 and NGC~362 had essentially indistinguishable ages which were
about 15\% smaller than the age of 47~Tuc. The parameters we have adopted for NGC~362 and 
NGC~2808
are given in Table~\ref{cluster_params}.

\begin{table}
\caption{Adopted parameters for NGC~362 and NGC~2808. For comparison we also give the data for NGC~104
(see Lebzelter et al.\,\cite{LW05}).}
\label{cluster_params}
\centering
\begin{tabular}{lccc}
\hline \hline
 & NGC 362 & NGC 2808 & NGC 104\\
\hline
$(m-M)_{V}$ & 14.80 & 15.56 & 13.50\\
$E(B-V)$    & 0.05  & 0.23  & 0.02\\
$M_{V,t}$\tablefootmark{a} & $-$8.43 & $-$9.39 & $-$9.42\\
age [yr] & 10$\times$10$^{9}$ & 10$\times$10$^{9}$ & 11$\times$10$^{9}$\\
Z & 0.001 & 0.001 & 0.004\\
\hline
\end{tabular}
\tablefoot{
\tablefoottext{a}{absolute visual magnitude from Harris\,(\cite{Harris96})}
}
\end{table}

\subsection{Previous searches for variable stars in our sample clusters}
As for most Milky Way globular clusters the searches for variable
stars predominantly focused on variables on the horizontal branch or
the main sequence and binaries. This has, in most cases, two
consequences: first, many AGB stars are already saturated on the deep
images required to reach the main sequence stars; and second, the
monitoring typically focuses on time scales of days or weeks rather
than the months or years which are needed to properly describe the
variability of red giants.  Therefore, as shown in Lebzelter \& Wood
(\cite{LW05}) for the well studied cluster 47~Tuc, a lot of long
period variables remained undiscovered.

However, in both NGC~362 and NGC~2808 a few long period variables have
been discovered.  For NGC~362 the Catalogue of Variable Stars in
Globular Clusters (Clement et al. \cite{Clement97}) lists two
candidates for LPVs, namely V2 (90 days period) and V16 (135
d). Concerning the period of V2 it has to be noted that an earlier
time series by Sawyer (\cite{SH31}) led to a period of 105\,d, while
the value of 90\,d listed in Clement et al. (\cite{Clement97}) stems
from a short time series from Eggen (\cite{Eggen72}). Eggen noted this
difference and attributed it to a possible period change that was also
suspected by Sawyer-Hogg, as she could not fit a couple of early
measurements of the star with the 105\,d period. V2 is also remarkable
due to its high Li abundance (Smith et al. \cite{Smith99}) indicating
an ongoing extra-mixing process in this star.  Data on the discovery
and period of V16 have been summarized by Lloyd-Evans (\cite{LE83}).
In the same paper it is shown that V16 has a high probability for
cluster membership according to radial velocity data. The most recent
detailed study has been published by Szekely et
al. (\cite{Szekely07}).  Beside V16 they detected four stars which
they classified as long period variables, two of them are probably
sources in the background. For one star, V17, they give a period of
approximately 69 d. The $K$-magnitude places the star on the RGB of
NGC~362. As Szekely et al. note, their fourth LPV V36 is likely a
cluster member according to the star's metallicity. V2 is not found in
their list of variables.

After the discovery of the red variable V1 in NGC~2808 in the 1960s,
Clement \& Hazen (\cite{CH89}) reported the finding of a second long
period variable, V11, and presented first data on the light changes of
these two stars. The variability of V1 has been confirmed by Corwin et
al. (\cite{corwin04}).  For none of the LPVs could a period be
derived.

\section{Observations and data analysis}
\subsection{Optical data}
The two clusters were monitored as part of a larger program searching for long period variables in
Galactic globular clusters. Results for 47~Tuc (Lebzelter \& Wood \cite{LW05}) and
NGC 1846 (Lebzelter \& Wood \cite{LW07}) have been published as the first papers from this 
monitoring program. 
The observations of NGC~2808 and NCG 362 were
obtained and analysed the same way. Thus we will give here only a brief summary and refer
to Lebzelter \& Wood (\cite{LW05}) for details.

The time series were obtained in two parts. The first part was
observed at Mount Stromlo using the 50inch telescope and the camera of
the MACHO experiment (Alcock et al. \cite{Macho92}). The MACHO camera
obtained two images in two broad band filters at the same time. These
passbands did not correspond to standard filters but the blue one was
similar to Johnson $V$. Observations started in May 2002, but came to an
abrupt stop after a few months when Mount Stromlo observatory was
destroyed by a bush fire.

We continued our monitoring program in late 2003 at CTIO's 1.3m
telescope operated by the SMARTS consortium. The instrument ANDICAM
was used for two monitoring runs in service mode from October 2003 to
January 2004 and from March 2005 to July 2005 (only NGC~2808),
respectively. Monitoring was done in Johnson $V$ and $I_C$.  

The $V$
and the $I_C$ data sets were analyzed separately. To detect variable
stars and to derive light curves we applied an image subtraction
technique using the ISIS 2.1 code developed by Alard (\cite{Alard00}).
Stellar fluxes on the reference frames were measured using the PSF
fitting software written by Ch. Alard for the DENIS project (see
Schuller et al. \cite{Schuller03}). A typical photometric accuracy
derived from 2 images obtained in the same night is 0$\fm$02 in $V$.

For both clusters we observed two fields with ANDICAM, each having
a field of view (FOV) of 6x6 arcmin. The central coordinates of the
two fields had an offset of approximately 3 arcminutes, i.e. the two
fields were overlapping by almost 50\,\%. The field of the MACHO camera
was considerably larger (even though we used only one segment out of 4), but no 
long period variables beyond the FOV covered by the ANDICAM images could be detected.
For NGC 362, Rosenberg et al. (\cite{rosenberg00}) give a core radius of 0.17 arcmin 
and a concentration
parameter log\,(r$_{tidal}$/r$_{core}$) of 1.94. For
NGC 2808 corresponding values of 0.26 arcmin and 1.77 are listed. Accordingly, the tidal
radius of both clusters is approximately 15 arcmin, and our observations, which
have been centered at the cluster centre, extend to only 20 to 25\,\% of the cluster's 
tidal radius. Since the stars of interest are relatively bright, they are
individually uncrowded and we expect the giant branch sample of detected stars
to be complete. In this region, the ratio of the cluster to the field star populations 
should be high.

To estimate the expected contamination by field stars in the direction of the two clusters
we used the TRILEGAL model (Girardi et al.
\cite{trilegal}) via its web interface\footnote{http://stev.oapd.inaf.it/cgi-bin/trilegal}. 
Standard settings were applied including a halo contribution. Within a field of 0.02 deg$^{2}$
(the size of the two ANDICAM fields in each cluster)
around the cluster centres the TRILEGAL output did not include any field stars that would by
chance fall onto the upper giant branch in a $V$ vs. $V-I$ colour-magnitude diagram. 

Among the variable stars detected, we selected the long period
variables based on the brightness ($V<14.5$ for NGC 362, and
$V<15$ for NGC 2808), timescale of the variation
(exceeding 30 days) and a total light amplitude in $V$ of at least 0.1
mag. Period search was done using Period98
(Sperl \cite{Sperl98}), a code which can compute a discrete Fourier transformation
in combination with a least-squares fitting of multiple frequencies on the data.  
A maximum of two periods was allowed for each
star. The period search was done independently on the light
curve segments from each of the several monitoring runs and then refined using the 
combined light curves. Stars with periods close to or exceeding the length
of one series of observations were analysed only on the basis of the
combined light curve.  We note that for the typical periods of tens of days found
for the variables, the light curves are well sampled so that aliasing should not be a
significant problem.  Due to the semi-regular nature of the variability in this kind of star,
any periodicity found represents only a snapshot of a
possibly more complex light curve, and some non-detected 
periodicities on much longer time scales may exist as well.

\subsection{Near-infrared data}
The ANDICAM instrument offers the possibility to get near-infrared
images at the same time as the optical image. During the time span
between October 2003 and January 2004 we obtained 12 epochs of
$K$-band data for NGC~2808 and 10 epochs for NGC~362, respectively.
However, the field-of-view is significantly smaller in the infrared than in the
optical, and the infrared (IR) field is a little bit off center. Therefore only
some of the variables identified in the optical could also be measured
in the infrared. For NGC~2808 we got time series for 11 stars and a
single measurement for two further objects. Only three variables could
be observed in NGC~362.

Photometry on the $K$-band images was performed in the same way as for
the visual data. The zero point of the magnitude scale was set using 4
to 8 stars on the frame that had 2MASS magnitudes.  Observed
magnitudes were converted to the 2MASS system by using the relation
given in Carpenter (\cite{Carpenter01}). All individual infrared
measurements for each variable (detected from visual data) were
combined to get an average $K$ magnitude. Using 2 nonvariable 2MASS
sources we estimate an accuracy of the mean $K$ brightness of
0.03\,mag.

\section{Results}
In both clusters a number of long period variables were found.  We
identified the variables in the 2MASS point source catalogue where
possible. With the exception of two stars (1 in NGC~362 and 1 in NGC
2808) close to the cluster centre we could in this way attribute
coordinates and 2MASS photometry data to our targets.

\subsection{NGC~362}
\begin{figure}
      \centering
      \resizebox{\hsize}{!}{\includegraphics{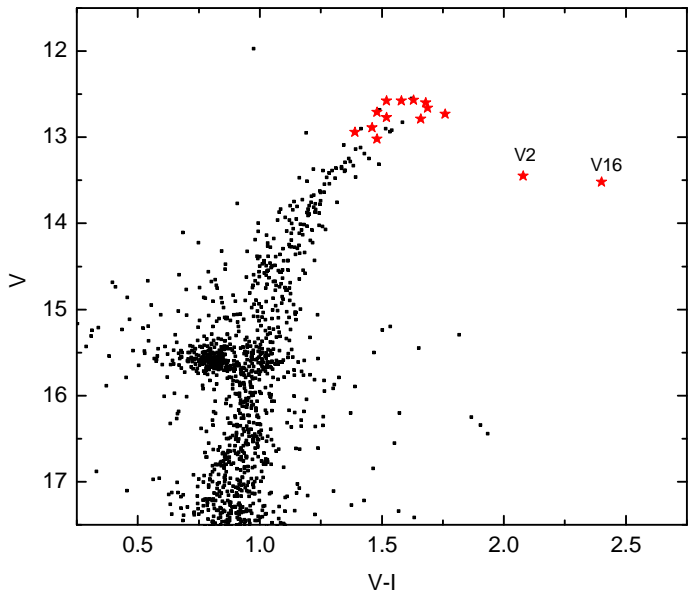}}
      \caption{CMD for NGC~362 derived from ANDICAM data. The long period
      variables are marked by red stars.}
      \label{fhd362}
   \end{figure}

\begin{figure*}
      \centering
      \includegraphics[width=17cm]{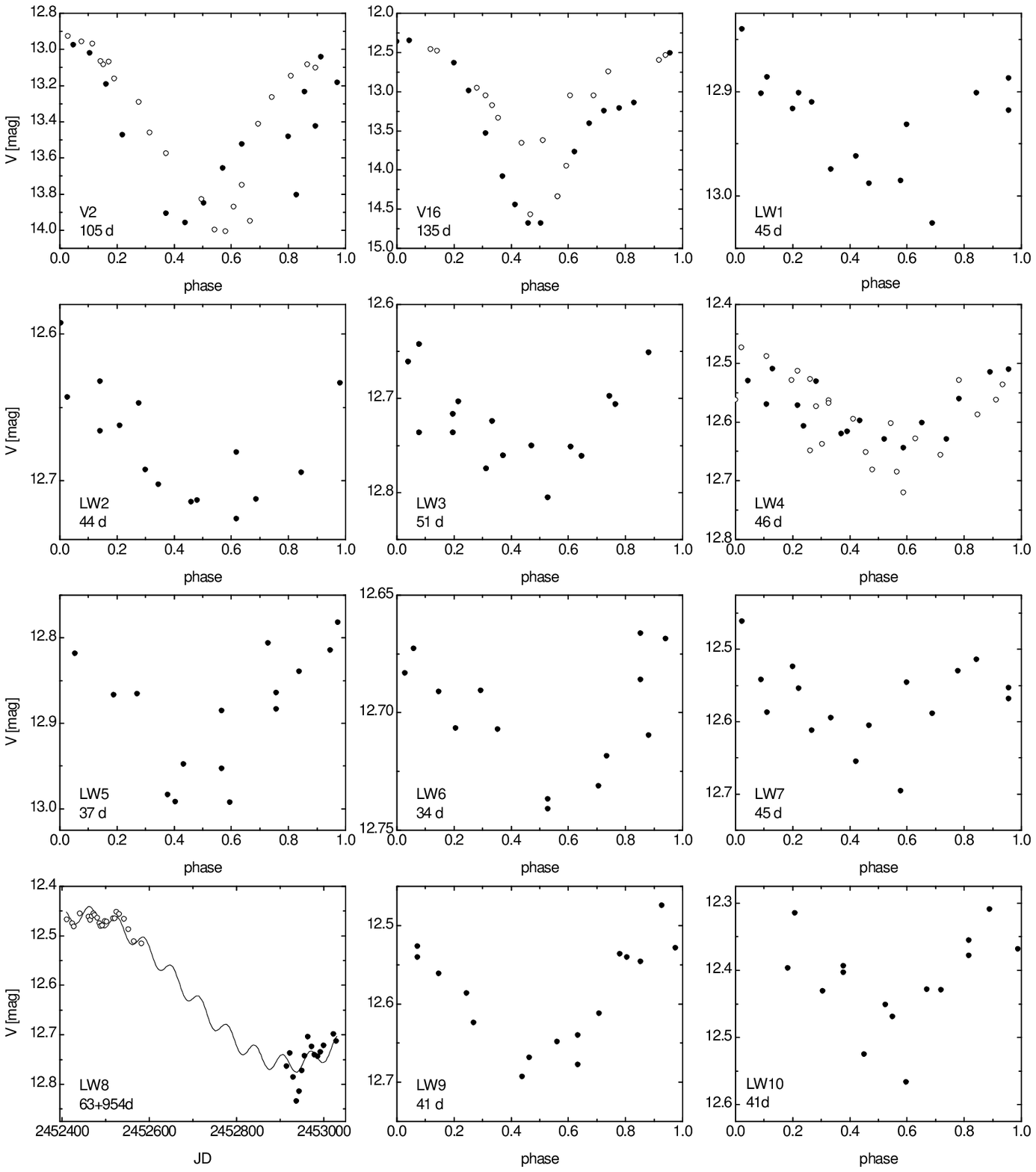}
      \caption{Phase diagrams of long period variables in our survey of NGC~362.
      In the lower left corner we show the light change of LW8 against time to illustrate the long-term
      variation of this star. The MACHO observations are shown as open circles while
      the ANDICAM observations are shown as filled circles}.
      \label{n362lc}
   \end{figure*}

Four known and seven new long period variables were found in NGC
362. For a large fraction of the variables in this cluster we had to
rely on the ANDICAM data only for determining the period, because for
the MACHO observations the central part of the cluster unfortunately
fell onto damaged areas of the detector for a significant fraction of
the time series.  The stars with ANDICAM data only are identified in
Table\,\ref{n362vars}, where coordinates, $V$, $I_C$, 2MASS magnitudes
and periods are listed. As in Lebzelter \& Wood (\cite{LW05}) we used the
variable star identifiers given by Clement et al. (\cite{Clement97})
where available. All other variables were named "LW" with some
number. Most of these variables have been detected for the first time,
see below for a few exceptions. The location of the variables in a
CMD of the cluster can be seen in
Fig.\,\ref{fhd362}. The phased light curves are presented in
Fig.\,\ref{n362lc}.

It is quite obvious that most stars show some kind of irregular
behaviour on top of the adopted periodic light change.  For some
targets a reasonable fit was only possible by adding a long secondary
period like in the case of LW8. However, an accurate period for these
long time variations could not be determined due to the limited length
of our time series.  In addition to the variables listed here we
noticed three other red stars that showed some kind of light
variability, but the variations did not follow any
(semi-)periodicity. These stars are listed at the bottom of Table
\ref{n362vars}, but were not investigated further.

Periods exist in the literature for two stars of our NGC~362
sample. As can be seen we confirm the period of 105\,d for V2
originally found by Sawyer (\cite{SH31}). It is not clear, if the star
is changing period from time to time or if the 90\,d period of Eggen
(\cite{Eggen72}) resulted from a short time irregularity in the light
curve\footnote{The time series used by Eggen (\cite{Eggen72}) to
derive the 89\,d period covered only one pulsation cycle.}. The
literature period for V16 is also found in our data.

Our star LW1 is identical with the star V36 of Szekely et
al. (\cite{Szekely07}). We confirm the suspicion that the star is a
long period variable. A second star from the list of Szekely et al.,
V56, is located very close to our variable LW6. However, these authors
classified it as an eclipsing binary. While we cannot completely
reject this hypothesis, we think that a classification as LPV seems to
be more probable.

Infrared data were added to Table \ref{n362vars} from the 2MASS
catalogue. As mentioned above our ANDICAM near-IR photometry
provided multiple $K$-magnitudes for only three sources from our list
(LW7, LW8 and LW9).  LW7 has a mean $K=8.72\pm0.05$ mag, for LW8 we find
$K=9.11\pm0.05$ mag. In both cases the scatter is above the expected
photometric accuracy, but no reasonable light curve could be
extracted. For LW9 we derived an average $K$ magnitude of
8.68\,mag. The star shows light variations with a $K$ amplitude of up
to 0.1\,mag, which are rather well synchronized with the observed
change in the visual. Our $K$-band photometry is in good agreement
with the 2MASS data.

\begin{table*}
\caption{Long period variables of NGC~362.}
\label{n362vars}
\centering
\begin{tabular}{lcccccccrl}
\hline \hline
ID & RA (2000) & Dec (2000) & $V$\tablefootmark{a} & $I_{C}$\tablefootmark{a} & $J$ & $H$ & $K$ & Period & Remark\\
   &           &           & [mag]   & [mag] & [mag] & [mag] & [mag] & [d] & \\
\hline
V2 & 01 03 21.9 & $-$70 54 20.1 & 13.45 & 11.37 & 9.67 & 8.99 & 8.76 & 105 & literature period 90 or 105 d\\
V16 & 01 03 15.0 & $-$70 50 32.3 & 13.52 & 11.12 & 9.10 & 8.44 & 8.17 & 135 & literature period 135 d\\
LW1 & 01 03 07.8 & $-$70 49 46.5 & 12.94 & 11.55 & 10.32 & 9.60 & 9.47 & 45 & only ANDICAM data, \\
    &            &             &       &       &       &      &      &    & V36 of Szekely et al. (2007)\\
LW2 & 01 03 10.7 & $-$70 50 54.0 & 12.66 & 10.97 & 9.85 & 8.97 & 8.75 & 44 & only ANDICAM data\\
LW3 & 01 03 13.6 & $-$70 50 37.0 & 12.73 & 10.97 & 9.73 & 8.89 & 8.61\rlap{:} & 51 & period from ANDICAM data\\
LW4 & 01 03 13.6\rlap{:} & $-$70 50 13.8\rlap{:} &12.60& 10.92 &      &      &       & 46 & no 2MASS identification\\
LW5 & 01 03 15.2 & $-$70 51 03.3 & 12.89 & 11.43 & 10.07 & 8.36 & 9.28\rlap{:} & 37 & only ANDICAM data\\
LW6 & 01 03 17.2 & $-$70 50 49.7 & 12.71 & 11.23 & 9.41 & 8.70 & 8.47 & 34 & only ANDICAM data,\\ 
    &            &             &       &       &      &      &      &    & probably V56 of Szekely et al.\\
LW7 & 01 03 19.0 & $-$70 50 51.4 & 12.58 & 11.06 & 9.79 & 8.93 & 8.75 & 45 & only ANDICAM data\\
LW8 & 01 03 33.0 & $-$70 49 37.2 & 12.77 & 11.25 & 10.06 & 9.30 & 9.10 & 63\rlap{:} & long time light trend\\
LW9 & 01 03 35.7 & $-$70 50 52.1 & 12.58 & 11.00 & 9.74 & 8.91 & 8.70 & 41 & only ANDICAM data\\
LW10 & 01 03 14.5 & $-$70 50 58.1 & 12.44\rlap{:} & 11.35\rlap{:} & 8.98\rlap{:} & 7.59 & 7.57\rlap{:} & 41 & only ANDICAM data\\
LW11 & 01 02 58.2 & $-$70 49 44.4 & 13.02 & 11.54 & 10.29 & 9.55 & 9.38 & ... & no period found\\
LW12 & 01 03 13.8 & $-$70 51 09.3 & 12.79 & 11.13 & 9.78\rlap{:} & 8.55 & 8.86 & ... & no period found\\
LW13 & 01 03 20.1 & $-$70 50 55.0 & 12.57 & 10.94 & 9.72 & 8.94 & 8.72 & ... & no period found\\

\hline
\end{tabular}
\tablefoot{A colon indicates a number with more uncertainty than usual.
\tablefoottext{a}{Mean values.} 
}
\end{table*}

Two investigations dedicated to cluster membership of stars in NGC~362
are found in the literature, namely the proper motion study by
Tucholke (\cite{Tucholke92}) and the radial velocity study of Fischer
et al. (\cite{Fischer93}). Proper motion data are also found in the
NOMAD database (Zacharias et al.  \cite{Zacharias05}). Five stars from
our sample have been measured by Tucholke: V2, LW8 and LW11 are listed
as proper motion members of the cluster. LW1 and LW9 have a low
probability for membership, but the proper motion data for theses two
stars come with a large error bar. LW8, LW10 and LW11 through LW13 are
cluster members according to their radial velocity. The
colour--magnitude diagram of the cluster (Fig. \ref{fhd362}) together with
the above mentioned test using the TRILEGAL stellar population model suggest
that all variables from our list are cluster members. While the cluster
is seen in projection on the outskirts of the Small Magellanic Cloud (SMC), any contamination
of the observed upper giant branch by extragalactic stars is expected
to be negligible due to the very low number density of SMC stars in that
colour-magnitude range (cf. Udalski et al. \cite{udalski08}). 

\subsection{NGC~2808}
\begin{figure}
      \centering
      \resizebox{\hsize}{!}{\includegraphics{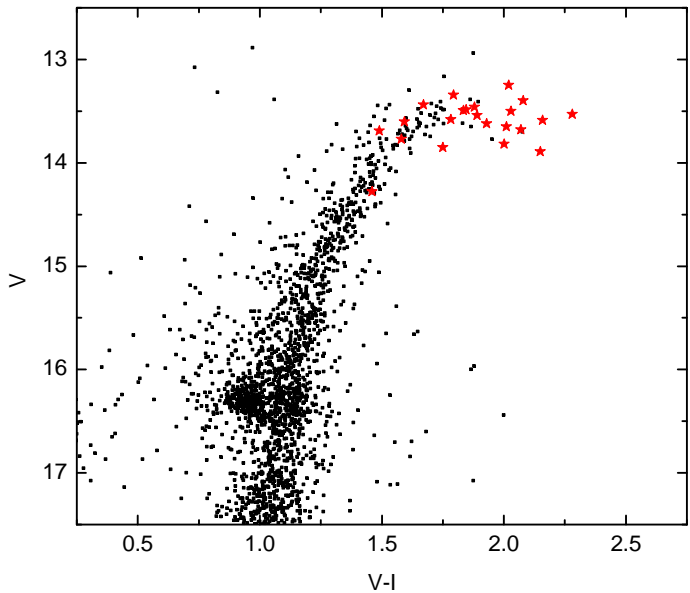}}
      \caption{CMD for NGC~2808 derived from ANDICAM data. The long period
      variables are marked by red stars.}
      \label{fhd2808}
   \end{figure}

\begin{figure*}
      \centering
      \includegraphics[width=17cm]{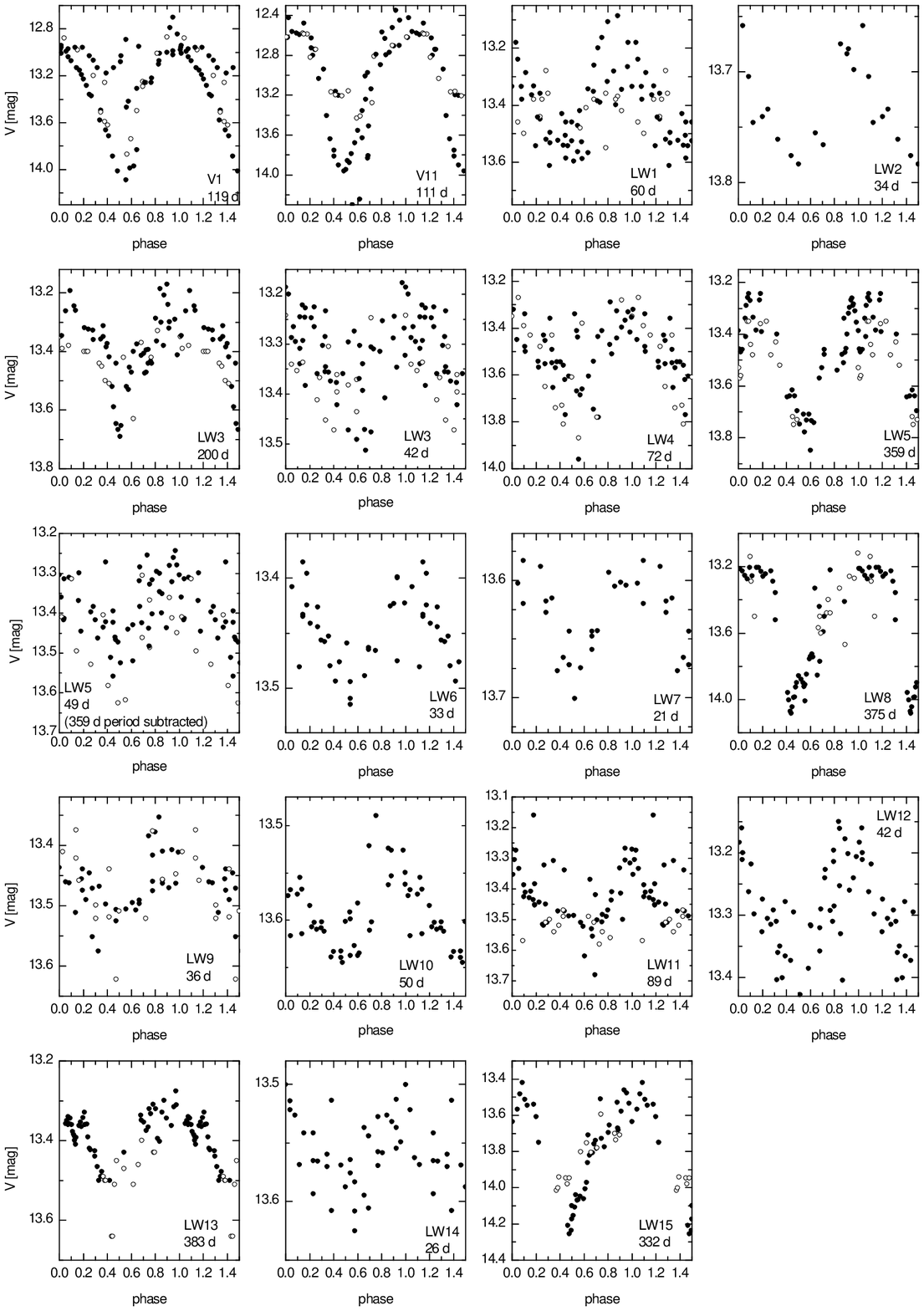}
     \caption{Phase diagrams of long period variables in our survey of NGC~2808.
      The MACHO observations are shown as open symbols while the ANDICAM
       observations are shown as filled symbols.} 
      \label{2808lcs}
    \end{figure*}

\begin{figure}
      \centering
      \resizebox{\hsize}{!}{\includegraphics{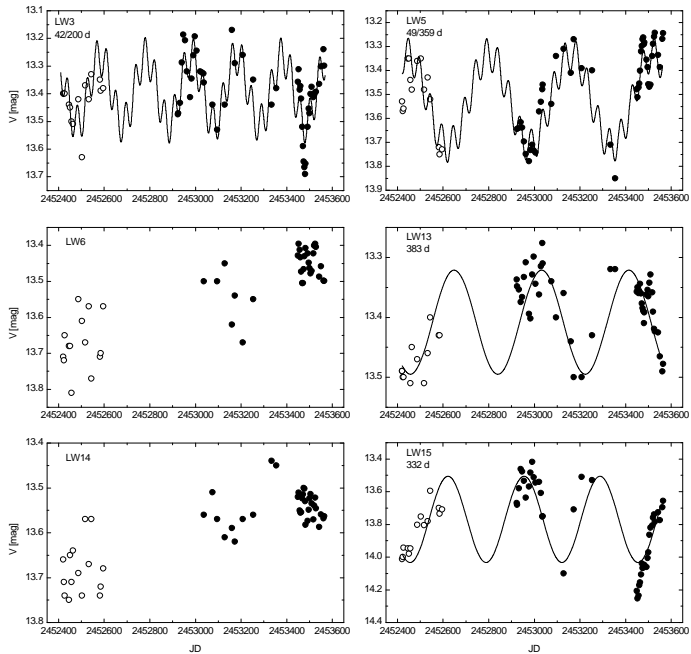}}
      \caption{Photometry of several variables in NGC~2808 plotted against Julian Date. 
      The corresponding Fourier fits are shown for stars with long periods.  For the two stars
      with long term trends in magnitude, only the observations are shown.  The MACHO observations
      are shown as open symbols while the ANDICAM observations are shown as filled symbols.}
      \label{2808lcs2}
   \end{figure}

Seventeen red variables of NGC~2808 showed a (semi-)periodic light
variation, including two variables known before (V1 and V11). 
Fig.\,\ref{fhd2808} shows the location of the variables in a
colour--magnitude diagram of the cluster. For 4
stars, all having a comparably short period, only the ANDICAM data,
which show a higher sampling in time, were used for the analysis.  The
corresponding phased light curves using the period with the largest
amplitude are shown in Fig.\,\ref{2808lcs}. 
In three cases (LW6, LW9 and LW14) we
subtracted a long time trend before producing the phase diagram.  For
LW2, ANDICAM and MACHO data were analysed independently and both gave
a very similar period. In the plot we show only one set of ANDICAM
data as there seems to be a phase shift between the older and the more
recent time series. 
In addition to the phase plots shown in Fig.\,\ref{2808lcs}, 
light curves plotted against time are shown in Fig.\,\ref{2808lcs2} for selected stars,
including those with long term trends, multiple periods or long secondary periods.
Beside the stars shown, we found 5 targets located
on the upper part of the giant branch, which clearly show variability
but without a clearly expressed periodicity. These stars are listed at
the bottom of Table \ref{n2808vars}.

As with the LPVs in NGC 362, irregularities in the light change can be
seen on top of the periodic variations. Some stars, e.g. LW5, show
clear signatures of an additional short time variation on top of the
longer period of 359 days. All reliably determined periods are listed
in Table \ref{n2808vars}. There is no source corresponding to LW15 in
the 2MASS catalogue. For this star the coordinates are estimated using
the SIMBAD/Aladin tool on the 2MASS $J$-band
image. 

\begin{figure}
      \centering
      \resizebox{\hsize}{!}{\includegraphics{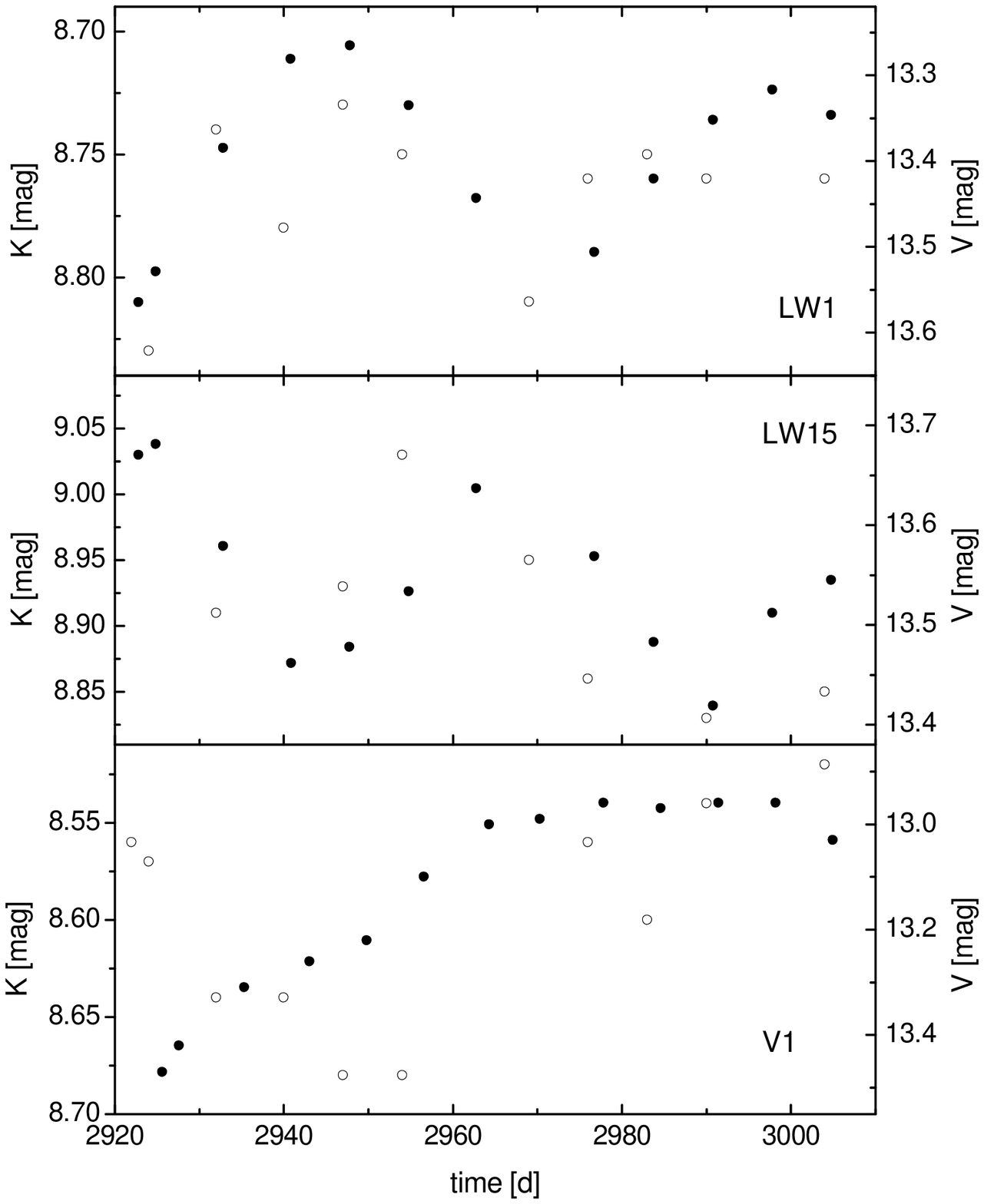}}
      \caption{Near infrared light curve (solid symbols, left y-axis) and visual light curve (open
      symbols, right y-axis) for LW1, LW15 and V1 of NGC~2808 (from top to bottom). Near infrared and
      visual data have been obtained at the same time.}
            \label{IR2808}
   \end{figure}

We also list in Table \ref{n2808vars} our own $K$-band photometry
(transfered to the 2MASS system) where available. The given
uncertainty of the $K$-band data includes both the measurement error
and the star's near-infrared variability during the four months of
monitoring.  For V11, LW9, LW18 and LW19 we have only a single epoch
measurement.  A well expressed near-infrared variability was detected
for three stars in our sample, V1, LW1 and LW15.  The light curves are
shown in Fig.\,\ref{IR2808} together with the visual data obtained at
the same time. It can be seen that for LW1 and LW15 the optical and
infrared lightcurves are nicely in phase, while for V1 there is a
large difference. The reason for that is not clear. The slightly
larger uncertainty of LW6 is most likely due to a higher photometric
error rather than due to variability. The star has another
near-infrared source nearby so that blending becomes a problem.  Note
that for LW15 we see the 332\,d period only as a general trend of the
average brightness in Fig.\,\ref{IR2808}. The star obviously shows
also a short period change of about 50 days length, which can be seen
as some scatter in Fig.\,\ref{2808lcs}.

For most sources we find a good agreement of our $K$-band measurement
with the 2MASS value taking into account the variability of the
objects. One exception is V1. Looking at the near infrared light curve
of V1 (Fig.\,\ref{IR2808}) the 2MASS value would suggest an amplitude
and thus a period much longer than the one measured from the visual
(the plotted range covers almost one complete light cycle). Therefore
we may speculate that the infrared light curve maybe expresses some
long time variation.  There is no indication for that in the visual
light curve. For two further stars, LW7 and LW19, the difference
between our value and the 2MASS data is also clearly exceeding the
typical uncertainty or variability scatter. We think the reason for
this is an error in the 2MASS value. For both stars the data point in
the 2MASS catalogue received a low quality flag, probably due to
crowding. For LW7, our value is the average of 11 measurements at
different epochs giving a standard deviation of $\sigma=$0.07 mag. The higher
value for $\sigma$ compared to some of the other low amplitude variables
in Table \ref{n2808vars} may be understood by the fact that it is the
faintest star in our sample.  For LW19 we have only a single
measurement.

\begin{table*}
\caption{Long period variables of NGC~2808.}
\label{n2808vars}
\centering
\begin{tabular}{lccccccccrl}
\hline \hline
ID & RA (2000) & Dec (2000) & $V$\tablefootmark{a} & $I_{C}$\tablefootmark{a} & $J$ & $H$ & $K$ & $K_{\rm andi}$ & Period & Remark\\
 & & & [mag] & [mag] & [mag] & [mag] & [mag] & [mag] & [d] & \\
\hline
V1 & 09 12 20.2 & $-$64 52 20.5 & 13.53 & 11.25 & 10.27 & 9.50 & 9.16 & 8.60$\pm$0.06 & 119 & \\
V11 & 09 12 06.7 & $-$64 52 40.3 & 13.25 & 11.23 & 9.77 & 9.04 & 8.82 & 8.92\tablefootmark{b} & 111 & \\
LW1 & 09 11 57.1 & $-$64 51 29.7 & 13.40 & 11.32 & 9.95 & 9.10 & 8.83 & 8.77$\pm$0.03 & 60 & \\
LW2 & 09 11 58.7 & $-$64 51 28.5 & 13.77 & 12.19 & 10.92 & 10.17 & 9.95 & 9.92$\pm$0.03 & 34 & ANDICAM only\\
LW3 & 09 11 59.2 & $-$64 52 59.2 & 13.50 & 11.47 & 10.02 & 9.07 & 8.85 & 8.83$\pm$0.03 & 42, 200 & 2 periods\\
LW4 & 09 12 01.4 & $-$64 51 37.4 & 13.65 & 11.64 & 10.24\rlap{:} & 9.06\rlap{:} & 9.24\rlap{:} & 9.16$\pm$0.07 & 72 &\\
LW5 & 09 12 01.7 & $-$64 50 33.2 & 13.59 & 11.43 & 10.18 & 9.26 & 9.02 & & 49, 359 & 2 periods\\
LW6 & 09 12 02.3 & $-$64 51 18.4 & 13.49 & 11.64 & 10.34 & 9.49 & 9.24 & 9.16$\pm$0.07 & 33 & long time trend\\
LW7 & 09 12 02.7 & $-$64 52 01.7 & 13.69 & 12.20 & 10.68\rlap{:} & 10.03\rlap{:} & 9.80\rlap{:} & 10.14$\pm$0.07 & 21 & ANDICAM only\\
LW8 & 09 12 04.3\rlap{:} & $-$64 51 41.7\rlap{:} & 13.68\rlap{:} & 11.61\rlap{:} & 9.86\rlap{:} & 8.96\rlap{:} & 8.75\rlap{:} & 8.87$\pm$0.12 & 375\rlap{:} & \\
LW9 & 09 12 06.7 & $-$64 52 18.1 & 13.54 & 11.65 & 10.23 & 9.35 & 9.10 & 9.08\tablefootmark{b}& 36 & ANDICAM only\\
LW10 & 09 12 08.5 & $-$64 51 58.4 & 13.62 & 11.69 & 10.34 & 9.43 & 9.25 & & 50 & long secondary P\\
LW11 & 09 12 11.3 & $-$64 52 42.2 & 13.52 & 11.66 & 10.29 & 9.35 & 9.17 & 9.07$\pm$0.05 & 89\rlap{:} & \\
LW12 & 09 12 14.5 & $-$64 53 26.7 & 13.34 & 11.55 & 10.18 & 9.30 & 9.07 & & 42 & ANDICAM only\\
LW13 & 09 12 16.6 & $-$64 52 02.8 & 13.46 & 11.58 & 10.23 & 9.35 & 9.10 & 9.05$\pm$0.05 & 383 & prob.~also short P\\
LW14 & 09 12 18.5 & $-$64 51 30.6 & 13.60 & 12.01 & 10.87 & 10.10 & 9.91 & 9.85$\pm$0.05 & 26 & ANDICAM only\\
     &            &               &       &       &       &       &      &               &    & long time trend\\
LW15 & (09 12 03.9) & ($-$64 51 54.2) & 13.89\rlap{:} & 11.74\rlap{:} &   &       &      & 8.93$\pm$0.09 & 332 & no 2MASS ID\\    
     &              &               &        &        &   &       &      &               &     &(crowded)\\
LW16 & 09 11 50.3 & $-$64 51 42.2 & 13.82 & 11.82 & 10.24 & 9.33 & 9.10 &  9.13$\pm0.03$ & ... & no period found\\
LW17 & 09 12 01.5 & $-$64 50 49.3 & 13.85 & 12.10 & 10.90\rlap{:} & 9.25\rlap{:} & 9.95\rlap{:} &  & ... & no period found\\
LW18 & 09 12 04.9 & $-$64 51 29.7 & 14.28 & 12.82 & 11.57 & 11.38 & 10.77 & 10.81\tablefootmark{b} & ...& no period found\\
LW19 & 09 12 05.9 & $-$64 52 08.1 & 13.58 & 11.80 & 10.38\rlap{:} & 9.51\rlap{:} & 9.24\rlap{:} & 9.39\tablefootmark{b} & ... & no period found\\
LW20 & 09 12 24.8 & $-$64 51 10.3 & 13.44 & 11.77 & 10.38 & 9.49 & 9.29 &    & ... & no period found\\
\hline
\end{tabular}\\
\tablefoot{A colon indicates a number with more uncertainty than usual.
\tablefoottext{a}{Mean values.}
\tablefoottext{b}{Only one single measurement in the $K$-band.}
}
\end{table*}

No proper motion study for individual stars in NGC~2808 could be found
in the literature. Membership information for our variables comes from
the study of red giant stars in this cluster by Cacciari et al.
(\cite{Cacciari04}). 
They find a mean radial velocity of approximately 101$\pm$10
km s$^{-1}$ for a sample of 132 objects they consider cluster members.
They also note that they found 5 additional objects with ''strongly discrepant''
velocities, and they presumed these to be field stars.   
The variables V1, LW1, LW3, LW5, LW6, LW9, LW10, LW12-14 and
LW17 have a measured radial velocity in Cacciari et al. (\cite{Cacciari04}). In all cases, 
there is a strong indication that these variables are cluster members,
with the variables having a mean radial velocity of 100.8$\pm$11.6 km s$^{-1}$.
There were no stars with strongly discrepant velocities.
Statistically, the results of  Cacciari et al. (\cite{Cacciari04}) suggest that the field 
population is about 3.6\% (5 out of 137 objects) which would indicate that
there could be about 0.8 field stars in our sample of 22 variables 
in NGC~2808.
As in the case of NGC~362 the CMD (Fig.
\ref{fhd2808}) also suggests that all the variables belong to this
cluster.

\section{The $\log P$--$K$ and $\log P$-$M_{\rm bol}$ diagrams}\label{sectpl}

As can be seen from the CMDs of the two clusters (Figs.\,\ref{fhd362}
and \ref{fhd2808}) all our variables are indeed long period variables
on the uppermost part of the giant branch and thus either AGB stars or
stars near the tip of the first giant branch.  Furthermore we may
notice that there is no non-variable star at the tip of the giant
branch. The same result was also found in 47~Tuc (Lebzelter \& Wood
\cite{LW05}).

Using these variables we constructed a $\log P$--$K$ diagram. Due to the
similarity in metallicity and age of the two clusters we decided to
combine the data into one diagram, but use separate symbols for the
two clusters. Only stars with a well defined periodicity and $K$
brightness were taken into account (no colon after the period or $K$
magnitude in Tables\,\ref{n362vars} or \ref{n2808vars}). 
The observed apparent $K$
magnitudes were corrected for reddening, using $E(B-V) = 0.05$ mag for NGC
362 and $E(B-V) = 0.23$ mag for NGC~2808, and distance, using
$(m-M)_{V} = 14.80$ and 15.56 mag,
respectively (we have taken these numbers from Piotto et al. 
\cite{Piotto02}).  The reddening law adopted (Rieke \& Lebofsky \cite{rl85}) 
gives $A_{\rm V} = 3.1 E(B-V)$
and $A_{\rm K} = 0.35 E(B-V)$.  In NGC~2808, we used the ANDICAM $K$ magnitude in 
preference to the 2MASS value.

\begin{figure}
      \centering
      \resizebox{\hsize}{!}{\includegraphics{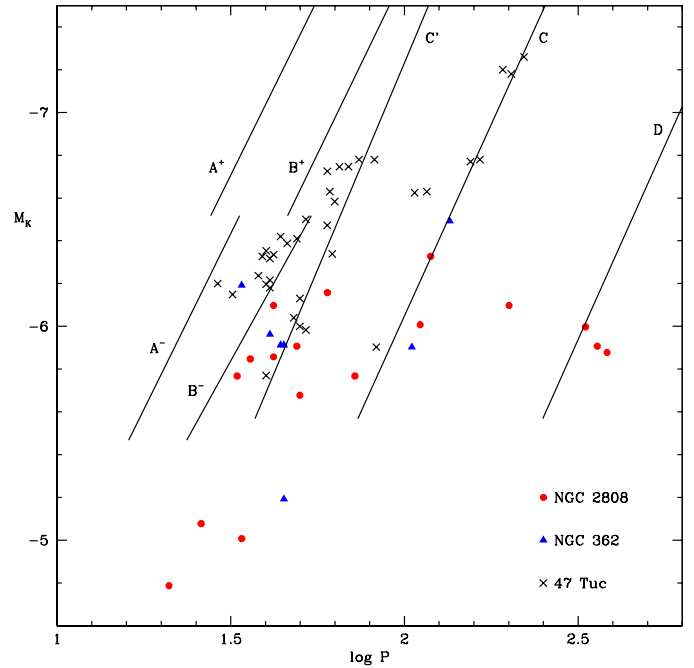}}
      \caption{LPVs in 47~Tuc, NGC~362 and NGC~2808 plotted in the $\log P$,$M_{\rm K}$ diagram
      along with the fits to the various LMC sequences given by Ita et al. (\cite{Ita04}).}
      \label{PLita}
  \end{figure}

In Fig.\,\ref{PLita} we present the combined $\log P$--$K$ diagram together
with the LMC relations given by Ita et al. (\cite{Ita04}).  The
variables found in 47~Tuc are also shown for comparison. NGC~362 and
NGC~2808 indeed seem to occupy a very similar area in the $\log P$--$K$
diagram supporting our approach to combine the two data sets in our
analysis. For an interpretation of this plot let us recall the
differences in the global parameters of NGC~362 and NGC~2808, 47~Tuc
and the LMC.  47~Tuc is more metal rich than NGC~362 and NGC~2808. As
summarized in the introduction section of this paper and in Lebzelter
\& Wood (\cite{LW05}), different relative ages have been found for
these three clusters and consequently they will have somewhat
different mass stars on the AGB.  The average metallicity of the LMC
(from which Ita et al. \cite{Ita04} derived their $\log P$--$K$ relations)
is higher than any of the clusters.  As the LMC is certainly not a
single age stellar population, the LPVs found are expected to
represent a wide range of stellar masses. It is not therefore clear
that the Ita et al. sequences should give a close fit to the NGC~362
and NGC~2808 data.

\begin{figure}
      \centering
      \resizebox{\hsize}{!}{\includegraphics{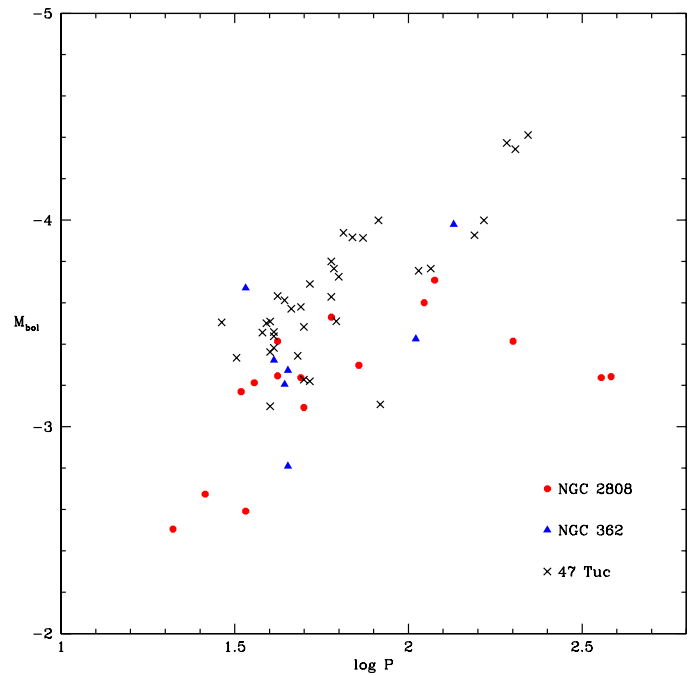}}
      \caption{The $\log P$,$M_{\rm bol}$ diagram for stars in 47~Tuc, NGC~362 and NGC~2808.}
      \label{mbollogp}
  \end{figure}

There is a clear difference both in the maximum and in the average
brightness of the LPVs in NGC~362 and NGC~2808 compared with 47
Tuc. The most luminous star (in the $K$-band) in 47~Tuc is almost
0.8\,mag brighter than the brightest LPV in NGC~362. Similarly, the
lowest luminosity LPVs found in 47~Tuc are brighter than the lowest
luminosity LPVs found in NGC~362 and NGC~2808.  Part of the reason for
the offset in $K$ is that the 47~Tuc stars are slightly cooler and
their bolometric corrections are larger than those of the NGC~362 and
NGC~2808 stars.  However, even in the $\log P$-$M_{\rm bol}$ diagram
(Fig.~\ref{mbollogp}), there still remains an offset of $\sim$0.5
magnitudes.  This difference is larger than the difference in RGB tip
luminosity which is $\sim$0.15 magnitudes brighter in 47~Tuc than in
NGC~362 and NGC~2808 according to the evolutionary tracks of Bertelli
et al. (\cite{Bertelli08}) for Z=0.004 and Z=0.001, respectively.

The same selection criteria were applied on all cluster data sets, and
the image quality was very similar so that variables should have been
detected to similar magnitude limits.  We therefore reach the
conclusion that the LPV distribution in NGC~362 and NGC~2808 is
$\sim$0.5 magnitudes fainter than in 47~Tuc.

In general, it is interesting to note that we found considerably fewer
LPVs in NGC~362 and NGC~2808 than in 47~Tuc.  Large amplitude
variables ($\Delta V > $2\,mag) are missing in NGC~362 and NGC~2808.
Frogel \& Elias (\cite{FE88}) suspected that there are no LPVs below a
metallicity of [Fe/H] $\approx$ $-$1. However, they defined LPVs on the
basis of the light amplitude only. The findings from our study shows
that this is not true if we consider low amplitude stars. Note also
that fundamental pulsators (sequence-C) are present in both clusters.
In this context it is also interesting to mention the detection
of two likely LPVs detected in the very metal poor cluster M15 ([Fe/H]\,=\,$-$2.3)
by McDonald et al. (\cite{MD10_2}). Their variations places them onto 
the pulsation sequence of the long secondary periods.

In 47~Tuc we saw a clear change of pulsation mode around
$M_{K}$\,=\,$-$6.8 mag. Similarly, in the two clusters studied here, overtone
pulsators extend up to a cutoff magnitude of $M_{K} \sim -$6.2 mag, which
is $\sim$0.6 mag fainter than in 47~Tuc.  There are two LPVs
brighter than this pulsating in the fundamental mode and, unlike in 47~Tuc,
there are a few stars below the maximum overtone magnitude that also
appear to belong to the fundamental mode sequence.

Fig.\,\ref{PLita} shows three stars in NGC~2808 that obviously 
belong to sequence-D.  No similar stars from 47 Tuc or NGC~362 are
seen in this figure.  However, this is largely the result of an
inability to determine reliable periods for the LSPs in these clusters.
In fact, Lebzelter \& Wood (\cite{LW05}) note that 12 of the red
variable in 47 Tuc show a long period but these periods could not be
determined from the relatively short duration of the observations.
The star LW3 in NGC~2808 which sits by itself at $\log P \sim 2.3$ 
and $M_{\rm K} \sim -6.1$ mag is discussed in Section 6.

\section{Observations of mass loss}

In our previous paper on 47~Tuc (Lebzelter \& Wood \cite{LW05}), we
showed that linear pulsation models including some parameterized mass
loss can describe the observed $\log P$--$K$ relations of single stellar
populations much better than models without mass loss.  These models
were then used to determine the mass loss rate on the giant
branch. Before we make an attempt to derive the mass loss on the red
giant branches of NGC~362 and NGC~2808 from their $\log P$--$K$ diagrams,
we want to briefly review previous studies that have investigated the
mass loss in these two clusters.

In NGC~362 the two previously
known variables V2 and V16 were found to show an infrared excess
suggesting emission from circumstellar dust (Ita et
al.\,\cite{Ita07}).
The most recent study of the dust mass loss from giants in NGC~362,
also studying an excess in the mid-infrared range, comes from Boyer et
al. (\cite{Boyer09}). 
We will not discuss here the numbers they give
for the mass loss rate since they depend on several assumptions, but
will focus instead on their identification of stars with a strong
infrared excess. Some of the sources they found in their work had
shown up previously in the investigations of Origlia et
al. (\cite{Origlia02}) and McDonald \& van Loon (\cite{MvL07}). Boyer
et al. identify 5 candidates for large mass loss rate and 5 candidates
for moderate mass loss rate from Spitzer photometry. Among the large
mass loss rate candidates they found two variables, namely V2 and
V16. Sloan et al. (\cite{Sloan10}) investigated these two stars
using IRS onboard the Spitzer space telescope, but found the mid-IR spectrum
featureless. They concluded that the infrared excess is caused by iron grains
(cf. McDonald et al. \cite{MD10}).
We identify another star of the list of Boyer et al., s07, as variable
LW3. Their source s05 seems to consist of two stars, one of which is
LW6.  They note that their fifth source with a large mass loss rate,
s08, is possibly a stellar blend. We found no variable star at that
position.  Thus, except for the uncertain case of s08, our data show
that all stars in NGC~362 with a strong mid-infrared excess and
deduced large mass loss rate are indeed long period variables. Among
the 5 stars with moderate mass loss rate from Boyer et
al. (\cite{Boyer09}) we can also identify 4 as variables. Most
interesting is the case of s03 which corresponds to our variable LW10,
the star detected from its near-infrared variability. The sources s01,
s04 and s09 correspond to LW9, LW7 and LW2, respectively. The fifth
star of Boyer et al. is outside the field of our study. This
comparison clearly shows the strong correlation between high mass loss
rates and variability.  

Mauas et al. (\cite{Mauas06}) derive a mass loss rate of a few
10$^{-9}$ $M_{\odot}$ on the upper part of the giant branch of NGC
2808 from the analysis of chromospheric lines.  H$\alpha$ line
components at an outflow velocity were found in a large sample of red
giants by Cacciari et al. (\cite{Cacciari04}). A Spitzer study of this
cluster has been obtained already (Fabbri et al. \cite{Fabbri08}), but
the results are not available yet.  From the horizontal branch
morphology of this cluster D'Antona \& Caloi (\cite{dAC08}) and
Dalessandro et al. (\cite{dal10}) derive an average mass lost before
the red clump phase of 0.15--0.20\,$M_{\odot}$ .

\section{Pulsation models and mass loss}

We have shown that modelling of the LPVs in 47~Tuc allows one to
derive the amount of mass lost on the RGB (Lebzelter \& Wood
\cite{LW05}).  It is our aim to carry out the same pulsation analysis
for NGC~362 and NGC~2808.  In addition, given the large range in the
helium abundance suspected in NGC~2808, we aim to see how the helium
abundance affects the LPV pulsation periods.  In particular, we aim to
see if the position on the period-luminosity diagram can give a clue to
the helium abundance.

Based on the summary of abundance measurements in the two clusters
detailed in the Introduction, we assume a metal abundance Z=0.001 for
both clusters.  Models have been made with helium mass fractions
Y=0.25, 0.3 and 0.4.  Note that these are, respectively, the estimated
helium mass fractions given by Dalessandro et al. (\cite{dal10}) for
red horizontal branch, blue horizontal branch and extreme horizontal
branch stars in NGC~2808.  Mass loss was added to some models assuming
a Reimers' Law (Reimers \cite{r75}). The Reimers mass loss rate was
multiplied by a factor $\eta$=0.4.  This factor was chosen as it leads to a
termination of the AGB near the maximum observed luminosity of the variable
stars in NGC~362 and NGC~2808.  It also closely reproduces the
masses estimated for the horizontal branch stars in the two clusters (see
below), and it is close to the value of 0.33 we found to apply in 47~Tuc.

\begin{figure}
      \centering
      \resizebox{\hsize}{!}{\includegraphics{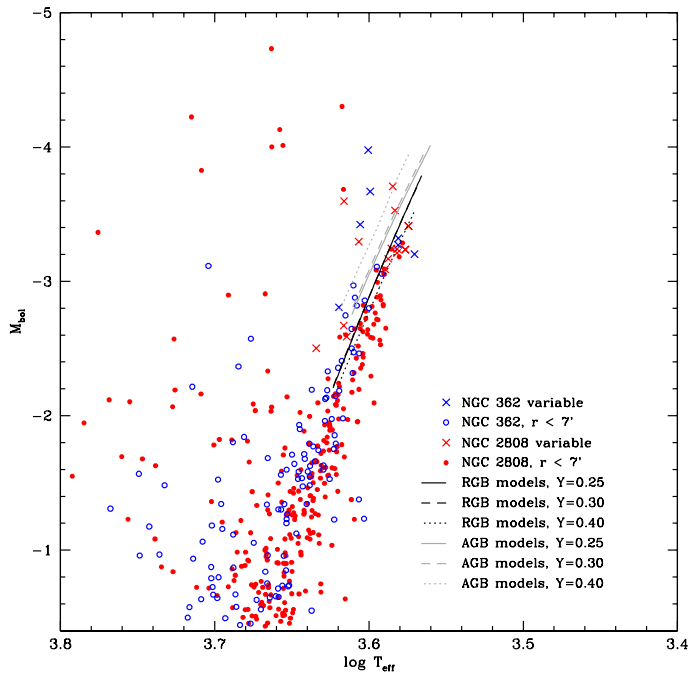}}
      \caption{Combined NGC~362 and NGC~2808 HR-diagram.  Luminosity
and $T_{\rm eff}$ were computed from $J$ and $K$ magnitudes (see text).
The locus of pulsation models without mass loss is shown for RGB and AGB stars with helium
abundance Y=0.25, 0.30 and 0.40.  All models have metallicity Z=0.001.
}
      \label{hrd_mconst}
\end{figure}

\begin{figure}
      \centering
      \resizebox{\hsize}{!}{\includegraphics{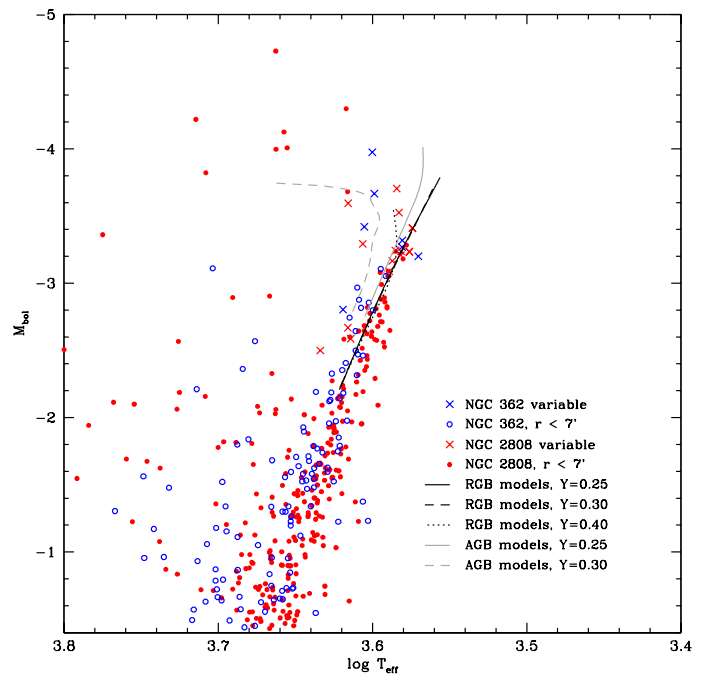}}
      \caption{Same as Fig.~\ref{hrd_mconst} but for models with mass loss.}
      \label{hrd_eta0.4}
   \end{figure}

In order to compute the declining
stellar mass along the RGB and AGB in the presence of mass loss,
evolution rates as a function of luminosity were obtained from the
evolution tracks of Bertelli et al (\cite{Bertelli08}).  $T_{\rm eff}$
in our calculations was forced to be the same as the mean observed
$T_{\rm eff}$ on the giant branch (see Fig.~\ref{hrd_mconst} or
Fig.~\ref{hrd_eta0.4}) so that the radius used in the Reimers' Law
formula was appropriate for NGC~362 and NGC~2808.

Following the discussion in the Introduction, we have assumed an age
of 10$\times 10^{9}$ years for both clusters and for each population
of different helium abundance.  The tracks and isochrones of Bertelli
et al.~(\cite{Bertelli08}) were used to obtain the initial mass
appropriate for giant branch stars of this age, Y and Z.  Initial
masses of 0.86, 0.79 and 0.65 $M_{\odot}$ were found for helium
abundances of 0.25, 0.3 and 0.4, respectively.

  \begin{figure}
      \centering
      \resizebox{\hsize}{!}{\includegraphics{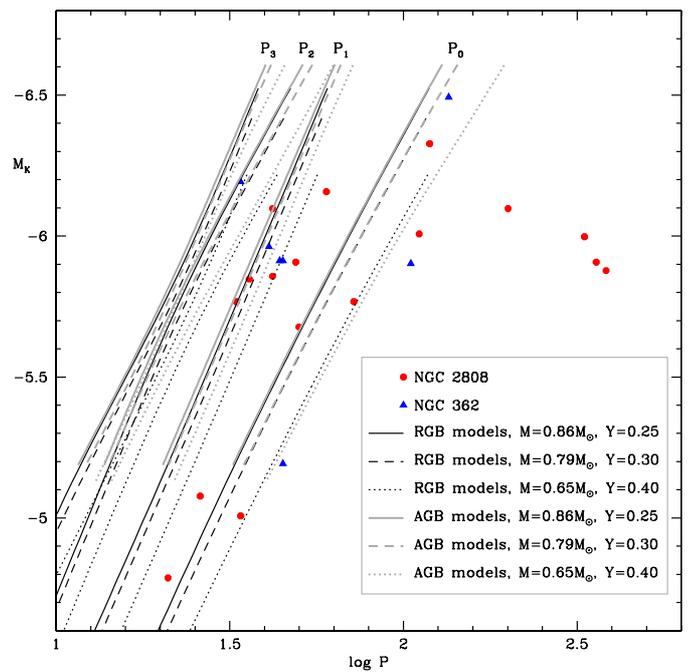}}
      \caption{The $\log P$--$K$ diagram for cluster variables and models without mass loss.
      The labels $P_{\rm 0}$, $P_{\rm 1}$, $P_{\rm 2}$ and  $P_{\rm 3}$ indicate radial 
      pulsation in the fundamental mode
      and the first, second and third overtone modes, respectively.
}
      \label{klogp_mconst}
  \end{figure}

Linear pulsation models were computed with the code described in
Lebzelter \& Wood (\cite{LW05}).  A mixing length of 1.7 pressure
scale heights was used in all models as this was found to reproduce
the observed $T_{\rm eff}$ of the giant branch well.

Models without mass loss are shown in the Hertzsprung-Russell-diagram in
Fig.~\ref{hrd_mconst}.  The $M_{\rm bol}$ and $T_{\rm eff}$ values of
the cluster stars shown in this figure were obtained from $K$ and
$J$-$K$ values from 2MASS for non-variable stars or from 
Tables~\ref{n362vars} and \ref{n2808vars} for variable stars.
Bolometric corrections for the $K$ magnitude
and the transformation from $J$-$K$ to $T_{\rm eff}$ given by Houdashelt
et al. (\cite{Houda00a}, \cite{Houda00b}) were used.  The distance 
moduli and reddenings we used are given in Section~\ref{sectpl}.  
In the HR-diagram, it can be seen that the models have
temperatures in the middle of the range of values of the variables.

The variables are shown in the $\log P$--$K$ diagram in
Fig.~\ref{klogp_mconst}.  For NGC~2808, where a wide range of helium
abundance is suspected, a model can be found to fit most variables.
However, in the main group of stars brighter than $M_{\rm K}=-$5.5,
nearly all the fundamental mode pulsators would require Y$ \sim $0.4,
which seems unlikely.  Also, the fundamental mode pulsator with $\log
P \sim 2.3$ has no explanation.  Since its amplitude is low, it is
unlikely that the period has changed as a result of large amplitude
pulsation, an effect demonstrated by Lebzelter \& Wood (\cite{LW05})
for the large amplitude Mira variables in 47~Tuc.  Finally, we note
that the 3 stars with $\log P \sim 2.5$ belong on sequence-D as shown
in Fig.~\ref{PLita}.  The origin of their variability is unknown
(e.g. Wood et al. \cite{wok04}; Nicholls et al. \cite{nic09}).

For NGC~362, where Y=0.25 should be appropriate, the models without
mass loss are a very poor fit to the observed $\log P$,$K$ values of
the variables.  We show below that a significant amount of mass loss on
the RGB can explain the periods of the LPVs in NGC~362 and
NGC~2808.

  \begin{figure}
      \centering
      \resizebox{\hsize}{!}{\includegraphics{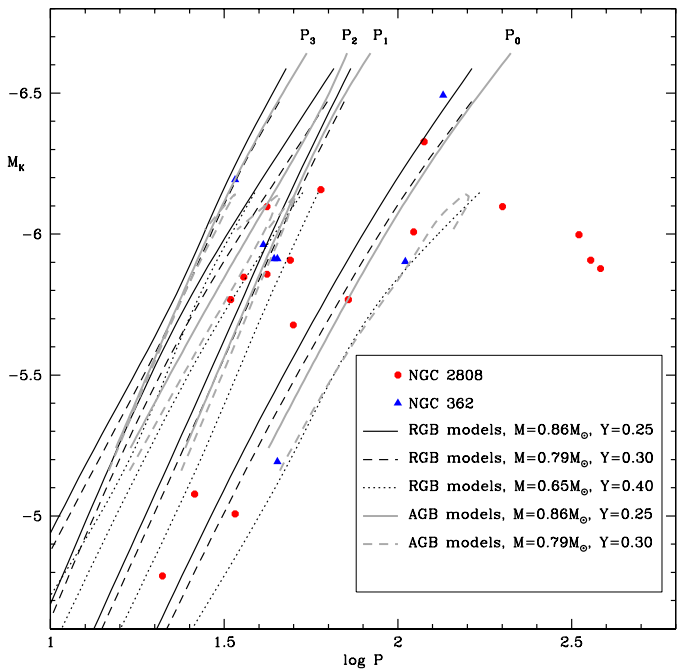}}
      \caption{Same as Fig.~\ref{klogp_mconst} but for models with mass loss.}
      \label{klogp_eta0.4}
  \end{figure}

Models with mass loss are shown in the Hertzsprung-Russell-diagram in
Fig.~\ref{hrd_eta0.4}.  With the adopted scaling parameter $\eta =
0.4$, it was found that the mass at the tip of the RGB (and hence on
the horizontal branch) was 0.693, 0.623 and 0.486 $M_{\odot}$ for
Y=0.25, 0.30 and 0.40, respectively.  This is in good agreement with
the values of 0.69, 0.565 and 0.479 $M_{\odot}$ found by Dalessandro
et al. (\cite{dal10}) for red, blue, and extreme horizontal branch 
stars in NGC~2808.  Note
that the AGB stars with Y=0.30 lose all their H-rich envelope at
$M_{\rm bol} \sim -3.7$ mag and then evolve away from the AGB to the blue.
Even more extreme are the stars with Y=0.4 which do not evolve up
the AGB at all since their H-burning shells consume the remaining
envelope hydrogen while on the horizontal branch. In general, the
$M_{\rm bol}$ and $T_{\rm eff}$ values of models fall in the middle of
the values obtained for the cluster stars.

The variables are shown in the $\log P$--$K$ diagram in
Fig.~\ref{klogp_eta0.4}.  In NGC~2808 most of the fundamental mode
pulsators can now be readily explained as RGB or AGB stars with Y=0.25
or RGB stars with Y=0.30.  The star LW3 in NGC~2808 with $\log P \sim 2.3$ now has an
interesting explanation as a fundamental mode pulsator at the very tip
of the RGB with a high helium abundance of Y\,$\sim$\,0.4, or perhaps
as an AGB star with Y\,$\sim$\,0.3.  The shorter period of LW3 (42 days)
is more consistent with overtone pulsation in the Y\,$\sim$\,0.3 models
than with the Y\,$\sim$\,0.4 models. 
We also note that two of the first overtone pulsators in NGC~2808 
have periods suggesting a helium abundance Y\,$\sim$\,0.4.  

If the explanation in terms of high helium abundance
is correct, it is completely independent evidence for a high helium
abundance in some stars in globular clusters.  It would be interesting
to see if other globular clusters which are estimated to have
populations of high helium abundance stars also have LPVs whose
periods can only be explained by a high helium abundance.  Note that
this test can not be used with RR Lyrae stars since the requirement
that the RR Lyraes lie in the instability strip confines their helium
abundance to a small range for a given cluster age.

In NGC~362 where Y = 0.25 should be appropriate, the models with mass
loss provide a better fit in most cases, although there are two
fundamental mode pulsators that appear to require some helium
enhancement (to Y\,$\sim$\,0.3 for AGB pulsation or Y\,$\sim$\,0.4
for RGB pulsation).

\section{Conclusions}

We have found and determined pulsation periods for LPVs in the 
Galactic globular clusters NGC~362 and NGC~2808.  By modelling the
pulsation of the LPVs, we have shown that mass loss is required on the
RGB at a rate that is close to that required to explain the morphology
of the horizontal branch.  The models show that helium abundance
variations of the size estimated from the multiple main-sequences in
NGC~2808 lead to a significant increase in the pulsation period of
LPVs, especially for the fundamental mode.  In NGC~2808, we have found
that the spread in fundamental mode periods suggests a spread in
helium abundance. One fundamental mode pulsator requires 
Y\,$\sim$\,0.3--0.4 if its period
is to be explained while the periods of two overtone pulsators are also best explained
with a helium abundance Y\,$\sim$\,0.4.  This is a completely independent line of evidence
for a high helium abundance in some stars in globular clusters.
Detection and analysis of LPVs in other clusters where helium
abundance variations are thought to occur could provide useful
confirmation of the high helium abundances.

\acknowledgements
The work of TL was funded by the Austrian Science Fund FWF project P20046-N16.  The
work of PRW was partially funded by Australian Research Council grant DP1095368.

{}

\end{document}